\documentclass[journal]{IEEEtran} 
\usepackage{ifpdf}
\usepackage{cite}
\usepackage{amsmath}
\usepackage{array}
\usepackage{float}
\usepackage{subfig}
\usepackage{cite}
\usepackage{graphicx}
\usepackage{amsmath}
\usepackage{epsfig}
\usepackage{epstopdf}
\usepackage{algorithm}
\usepackage{algorithmicx}
\usepackage{algpseudocode}
\usepackage{blkarray}
\usepackage[table,xcdraw]{xcolor}
\usepackage{float}
\usepackage{amsfonts}
\usepackage{mathrsfs}
\ifCLASSINFOpdf
\else
\fi

\begin{document}
\title{Role Switching and Power Allocation Technique for Mobile Users in NOMA}

\author{Irfan Azam,~\IEEEmembership{Student Member, IEEE}, Muhammad Basit Shahab
, and Soo Young Shin,~\IEEEmembership{Senior Member, IEEE}
\thanks{Irfan Azam and Soo Young Shin are with WENS Lab, Department of IT Convergence Engineering, Kumoh National Institute of Technology, Republic of Korea, (Email: irfanazam@kumoh.ac.kr, wdragon@kumoh.ac.kr)}
\thanks{Muhammad Basit Shahab is with School of Electrical Engineering and Computing, The University of Newcastle, Australia, (Email: basit.shahab@newcastle.edu.au)}
}


\maketitle
\begin{abstract}
In this letter, role switching and power allocation schemes are proposed to tackle user mobility in non-orthogonal multiple access (NOMA) systems. When cell center user (CCU) and cell edge user (CEU) come very close or even cross each other in NOMA pairing, channel gains of paired users violate the basic NOMA conditions. This article refers to such condition as NOMA principle violation problem (NPVP). To solve this NPVP, optimized power role switching-NOMA (OPRS-NOMA) technique is proposed. Role switching technique is used where roles of mobile users are switched on the basis of their channel gains. Furthermore, a power allocation scheme based on bisection search power optimization is presented to maximize the average sum capacity of mobile NOMA users. Random way point mobility model is considered for user mobility. Individual and sum capacity are used for performance evaluation. Simulation results show that OPRS-NOMA outperforms the conventional NOMA and orthogonal multiple access (OMA).
\end{abstract}
\IEEEpeerreviewmaketitle
\section{Introduction}
\IEEEPARstart{T}{o} meet the capacity and connectivity demands of a large number of expected devices, non-orthogonal multiple access (NOMA) has been adopted as multi-user superposition transmission (MUST) by the 3GPP release-16 standards for 5G \cite{1}. In NOMA, multiple users are served over the same radio resource block (RB), where base station (BS) allocates different power levels to these paired/grouped users to facilitate efficient data recovery through successive interference cancellation (SIC).

The existing works in NOMA \cite{2,3}, mostly focus on the capacity maximization of NOMA system with static users. However, if paired users are mobile, they may violate the basic NOMA channel gain condition. NOMA works only when a cell center user (CCU), i.e., near user has a better channel gain than the paired cell edge user (CEU), i.e., far user. Let $|h_1|^2$ and $|h_2|^2$ be the channel gains of CCU and CEU from the BS, respectively. NOMA can work normally when $|h_1|^2 \gg |h_2|^2$. However, if users are mobile, they can come very close to each other or cross each other so that the channel gains could be $|h_1|^2 \approx |h_2|^2$ or $|h_1|^2 < |h_2|^2$. In this paper, this situation is called as NOMA principle violation problem (NPVP). To the best knowledge of authors, research on user pairing and power allocation in NOMA with user mobility is still lacking. Therefore, the work in this article focuses specifically on mobility issues in NOMA.

To resolve NPVP in mobile environment, an optimized power role switching-NOMA (OPRS-NOMA) technique is proposed in this paper. The role switching technique between CCU and CEU is firstly introduced, which changes the roles of CCU and CEU, when their channel gain ordering is inverted. Furthermore, an optimal power allocation technique is presented to maximize sum capacity of CCU and CEU. To validate the effectiveness of proposed techniques, random way point (RWP) mobility model is used for NOMA users.

\section{System Model and Problem Formulation}
\subsection{NOMA Protocol}
In NOMA, multiple paired users can be simultaneously served over the same radio RB. Consider two mobile NOMA users; mobile $\text{UE}_1$ ($\text{MUE}_1$) and mobile $\text{UE}_2$ ($\text{MUE}_2$), with channel gains $|h_1|^2$ and $|h_2|^2$, where $|h_1|^2 \geq |h_2|^2$. Channel $|h_i|$ is considered to be independent Rayleigh flat fading with channel coefficient $h_i \sim CN$ $(0,\lambda_i=d_i^{-v})$ having mean 0 and variance $\lambda_i$ for the $\text{BS}-\text{MUE}_i$ link, where $d_i$ is the $\text{BS}-\text{MUE}_i$ distance, and v is the path loss exponent.  For simplicity, single-input and single-output (SISO) antenna configuration is considered. The distances from BS to $\text{MUE}_1$ and $\text{MUE}_2$ are $d_1$ and $d_2$ respectively. The PA factors of both users are $\rho_1$ and $\rho_2$, where $\rho_1$ + $\rho_2=1$, and $\rho_2 > \rho_1$. Considering $P_t$ as the total transmit power of BS, the individual user capacity for $\text{MUE}_1$ (CCU) and $\text{MUE}_2$ (CEU) can be written as
%
\begin{equation}
\label{eq01}
R_1 = \log_2 \left( 1 + \frac{\rho_1 P_t |h_1|^2}{N_o}  \right)
\end{equation}
\begin{equation}
\label{eq02}
R_2 = \log_2 \left( 1+\frac{\rho_2 P_t |h_2|^2 }{\rho_1 P_t |h_2|^2 + N_o} \right)
\end{equation}
\par
where $N_o$ represents variance of the additive white Gaussian noise (AWGN).
Correspondingly, pair sum capacity (PSC) of the users can be calculated as
\\
$R_{\text{sum}}=R_1+R_2$
\\
\begin{equation}
\label{eq03}
=\log_2 \left( 1 + \frac{\rho_1 P_t |h_1|^2}{N_o}  \right) +\log_2 \left( 1+\frac{\rho_2 P_t |h_2|^2 }{\rho_1 P_t |h_2|^2 + N_o} \right).
\end{equation}

\begin{figure}[!t]
\centering
\includegraphics[scale=0.3]{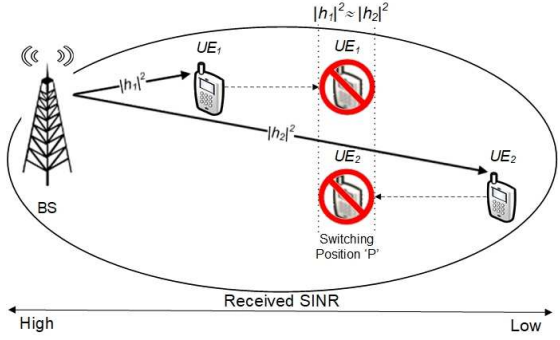}
\caption{NOMA violation limit at switching position 'P', after which,  mobile users roles should be switched.}
\label{Figure:0}
\end{figure}

\subsection{Mobility Model}
In the proposed system, RWP mobility model is considered for user mobility, while BS is fixed at the center of a cellular area. MUEs change their position from point $p_o$ to $p_n$ at each time instant $t_s$. The position of a $\text{MUE}_i$ is uniformly selected within the specified area and its velocity is also selected from the minimum and the maximum velocity interval [$vt_{\min}, vt_{\max}$]. Accordingly, the BS to $\text{MUE}_i$ distance $d_i$ is calculated at each new position $p_n$. A 2D model is considered with distance as a random variable and generalized probability density function (PDF) given as \cite{4}
\begin{equation}
\label{eq04}
f_d(d) = \sum_{i=1}^n \beta_i \frac{d^{\beta_i}} {D^{\beta_i+1}},~~~~0 \leq r \leq D,
\end{equation}
where the mobility parameters are $n=3$, $B_i = \left( \frac{1}{73} \right) . \left[ 324, -420, 96 \right]$, and $\beta_i = \left[1,3,5\right]$, for 2D topology.
\begin{figure*}[!t]
    \centering
    \null\hfill
    \subfloat[Normalized distance of $\text{MUE}_1$ and $\text{MUE}_2$ from the BS and the positions where the roles should be switched.] {\includegraphics[width=0.4\textwidth]{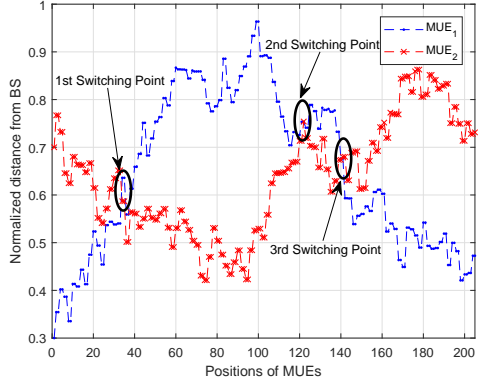}}
    \hfill
    \subfloat[Individual capacity of $\text{MUE}_1$ and $\text{MUE}_2$ in conventional NOMA under RWP mobility model without switching.]{\includegraphics[width=0.4\textwidth]{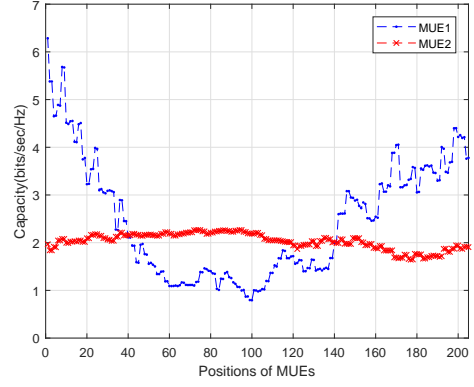}}
    \hfill\null
    \caption{\label{Figure:1}
    RWP based mobility illustration of $\text{MUE}_1$, $\text{MUE}_2$, and their corresponding capacity results without role switching.}
\end{figure*}

\begin{algorithm}
	\caption{\textbf{O}ptimized \textbf{P}ower \textbf{R}ole \textbf{S}witching (OPRS-NOMA)}
	
	\hspace*{\algorithmicindent} \textbf{Input:} \\
    \hspace*{\algorithmicindent} Speed interval: $vt_{\min}$ to $vt_{\max}$. \\
	\hspace*{\algorithmicindent} $X$ and $Y$ position interval: $(x_{\min},y_{\min})$ to $(x_{\max},y_{\max})$. \\
	\hspace*{\algorithmicindent} Direction angle interval: $-\pi$ to $\pi$.\\
	\hspace*{\algorithmicindent} Number of users: $N$.\\
	\hspace*{\algorithmicindent} Distance between BS and $\text{MUE}_i$ ($D_{\text{BS}-\text{MUE}_i}$): $1 \times N$. \\
    \hspace*{\algorithmicindent} \textbf{Output:} \\
    \hspace*{\algorithmicindent} $R_1$ and $R_2$ \\
	\hspace*{\algorithmicindent} \textbf{\emph{Notation}}: Subscripts 1 and 2 represents near and far \\
	\hspace*{\algorithmicindent} \qquad \qquad~users respectively. \\
	\hspace*{\algorithmicindent} \textbf{Initialization:} \\
    \hspace*{\algorithmicindent} Channel difference threshold $CH_{th}$.	\\
    \hspace*{\algorithmicindent} Power difference threshold $\rho_{th}$.
	\begin{algorithmic}[1]
		\State \textbf{Generate RWP mobility} (Input)
		\State ~~~~~~~~~~~~\{\textbf{return} position P\}
		\For {each position P of $\text{MUE}s$}
		\State Calculate Distances: $D_{\text{BS}-\text{MUE}}$
		\State Compute channel matrix \textbf{H} from $D_{\text{BS}-\text{MUE}}$
        \State Calculate channel gains: $|h_1|^2, |h_2|^2$
		\If {$|h_1|^2 > |h_2|^2 ~\textbf{AND}~ |h_1|^2 - |h_2|^2 > CH_{th}$}
		\State Apply BSPO without power threshold $\rho_{th}$
		\State \{\textbf{return} ($\rho_1,\rho_2$)\}
		\ElsIf {$|h_1|^2 > |h_2|^2 ~\textbf{AND}~ |h_1|^2 - |h_2|^2 \leq CH_{th}$}
		\State Apply BSPO with power threshold $\rho_{th}$
		\State \{\textbf{return} ($\rho_1,\rho_2$)\}
		\ElsIf {$|h_1|^2 < |h_2|^2 ~\textbf{AND}~ |h_2|^2 - |h_1|^2 \leq CH_{th}$}
		\State Apply Role Switching $\qquad \triangleright~ \footnotesize \text{users ordering switched}$
		\State Apply BSPO with power threshold $\rho_{th}$
		\State \{\textbf{return} ($\rho_1,\rho_2$)\}
        \Else  \qquad $ \triangleright~|h_1|^2 < |h_2|^2 ~\textbf{AND}~ |h_2|^2 - |h_1|^2 > CH_{th}$
		\State Apply Role Switching $\qquad \triangleright~ \footnotesize \text{users ordering switched}$
		\State Apply BSPO without power threshold $\rho_{th}$
		\State \{\textbf{return} ($\rho_1,\rho_2$)\}
		\EndIf
		\State Calculate eq. (\ref{eq01}) and eq. (\ref{eq02})
		\EndFor
	\end{algorithmic}
\end{algorithm}
\subsection{Problem Formulation}
The problem at hand is the role switching of MUEs to keep the NOMA system working under mobility, and optimizing their PA factors to maximize PSC of the pair. The overall optimization problem can be expressed as
\begin{subequations} \label{e_df_minmax_II}
\small \begin{alignat}{2}
& \max\limits_{\rho_1,\rho_2}
& & (R_1+R_2) \label{eq:5a}\\ 
& ~ \text{s.~t.~}& & \rho_2 \geq \rho_1, ~ \rho_1+\rho_2 = 1,~ \text{if}~|h_1|^2-|h_2|^2 \geq CH_{th} \label{eq:5b}\\
& & &  \rho_2 - \rho_1 \geq \rho_{th},~ \rho_1+\rho_2 = 1,~\text{if}~|h_1|^2-|h_2|^2 \leq CH_{th} \label{eq:5c} \\
& & & x_{\min} \leq X_P \leq x_{\max}, y_{\min} \leq Y_P \leq y_{\max}. \label{eq:5d}
\end{alignat}
\end{subequations}
The optimization function in (\ref{eq:5a}) maximizes the PSC, (\ref{eq:5b}) is the general power allocation rule with distant users, (\ref{eq:5c}) provides a minimum power difference constraint between users if their channel gains difference is less than a threshold (users close to each other), and (\ref{eq:5d}) represents the minimum/maximum limits for user mobility.

\begin{figure*}[!t]
	\centering
	\null\hfill
	\subfloat[$\text{MUE}_1$, $\text{MUE}_2$ capacity; role switching with fixed power.] {\includegraphics[width=0.4\textwidth]{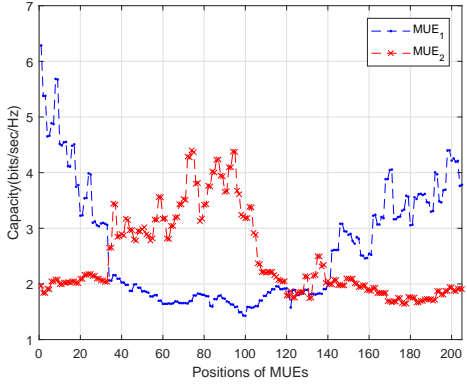}}
	\hfill
	\subfloat[$\text{MUE}_1$, $\text{MUE}_2$ capacity; role switching with optimal power.]{\includegraphics[width=0.4\textwidth]{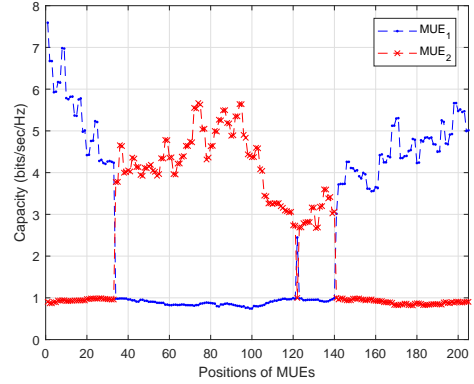}}
	\hfill\null
	\caption{\label{Figure:2}
		Comparison of fixed power role switching and optimized power role switching.}
\end{figure*}

\section{Role Switching and Power Optimization}
The proposed work address the NPVP in NOMA under user mobility by considering the variable channel gains of users that serve as a baseline of NOMA pairing. In order to show the working principle of the proposed scheme, consider that the two $\text{MUEs}$ ($\text{MUE}_1, \text{MUE}_2$), following RWP mobility model, come very close or even cross each other multiple times. Every time the MUEs come very close to each other, their channel conditions at the switching position $\text{P}$ become approximately similar. Conventional NOMA PA schemes that focus on sum capacity maximization may assign very close PA factors to both users, without caring about the data recovery problems at the user ends. Moreover, if the users cross each other, then their roles as near/far users and the associated PA ordering both need to be switched. In situations when users come very close i.e. $|h_1|^2-|h_2|^2 \leq CH_{th}$ to each other, the OPRS-NOMA scheme ensures the power difference between paired users to be greater than a predefined threshold $\rho_{th}$, i.e.,
\begin{align}
\label{eq06}
&{} \rho_2-\rho_1 \geq \rho_{th}.
\end{align}
Conventionally, PA of the users should not be equal, $\rho_1$ should be smaller than 0.5, and $\rho_2$ larger than 0.5. When both users are close to each other, and CEU has small target rate, then in order to maximize PSC, BS can allocate $\rho_1=0.49$ and $\rho_2=0.51$, as giving maximum allowed power to CCU improves the PSC. This is similar to bisection search power optimization (BSPO [2]). However, such close PA factors of the two users can significantly degrade their data recovery process; a predefined threshold based PA can resolve the issue.
\par
The proposed OPRS-NOMA works by checking the channel conditions of the paired users at each new location. In case their channel gain ordering is the same as previous, the algorithm only focuses on the power optimization part; BSPO or BSPO with threshold. However, if the channel gain ordering of the users change, then the algorithm changes their roles as near/far users (and inverts their power order) followed by power optimization using BSPO or BSPO with threshold. This can be noticed in the if-else conditions in Algorithm 1, which are based on the channel gains of paired users. The first two conditions refer to the case where channel ordering of the paired MUEs does not change, and therefore no role switching is needed. The difference between these two conditions is whether the channel gain difference of MUEs is larger or smaller than a threshold $CH_{th}$ i.e., the MUEs are far away from each other or closer. Large channel gain difference corresponds to BSPO, while BSPO with threshold $\rho_{th}$ is used otherwise. The other two conditions refer to the case where channel ordering of the users changes (near and far users cross each other), which requires both role switching and PA. The difference between both these conditions is also in terms of the channel gain difference. Both conditions refer to the case where users ordering change and their roles and PA need to be switched. In case the channel gains are still close to each other (just after crossing each other), then role switching and BSPO with threshold is used. Otherwise, role switching with conventional BSPO is used.
\par
In short, the OPRS-NOMA overcomes the NPVP of conventional NOMA by checking the channel condition of both users, which should be greater then the predefined channel threshold $CH_{th}$. Secondly, the minimum power threshold $\rho_{th}$ condition is also checked if the channel gains are less than channel threshold $CH_{th}$. Finally, the OPRS-NOMA switches the roles and performs power optimization if required on the basis of channel difference by using the conventional BSPO technique discussed in \cite{2}.
\begin{figure*}[t]
    \centering
    \null\hfill
    \subfloat[PSC gain before and after switching with fixed and optimized power in OMA and NOMA systems.]{\includegraphics[width=0.4\textwidth]{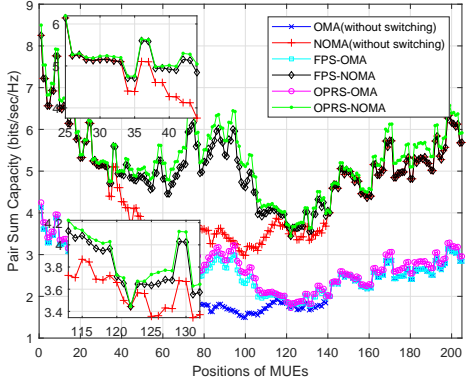}}
    \hfill
    \subfloat[PSC gain of the OPRS-NOMA in the interval between two switchings.] {\includegraphics[width=0.4\textwidth]{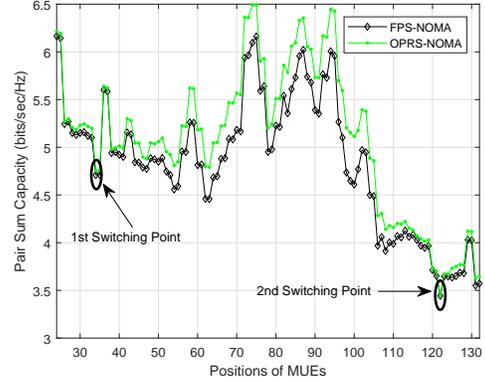}}
    \hfill\null
    \caption{\label{Figure:3}
    Detailed comparison in terms of PSC of the proposed role switching techniques in OMA and NOMA systems and the gain achieved by the OPRS-NOMA during role switching.}
\end{figure*}
\section{Simulation Results and Discussion}
Consider two MUEs ($\text{MUE}_1$, $\text{MUE}_2$) following RWP model in a downlink NOMA system that change their positions, speed and direction at each time instant $t_s$. Rest of the parameters are set as target rate $T_R=1$ bit/s/Hz, bandwidth $B=1$ Hz, signal-to-noise ratio $\text{SNR}=5$dB, and the BS to MUEs distances normalized to 1. Performance is evaluated in terms of individual and sum capacity (PSC) of paired users.
\par
In Fig. \ref{Figure:1}(a), normalized BS to MUEs distances are shown at different time instant $t_s$ and the users 1st switching point at $\text{P}=34$ is shown, where the channel condition of the $\text{MUE}_1$ becomes worse than the $\text{MUE}_2$ thereby violating the NOMA principle. It can be seen in Fig. \ref{Figure:1}(b) that without role switching, capacity of $\text{MUE}_2$ is still not increased as its channel condition becomes better than $\text{MUE}_1$. The individual user capacity results are shown in Fig. \ref{Figure:2}a after applying role switching with fixed power to show the benefits. Significant capacity gain of near and far users can be noticed after roles are switched at $\text{P}=34$. Furthermore, the capacity is further maximized by applying the BSPO \cite{2} with role switching in Fig. \ref{Figure:2}(b).
\par
\begin{figure}
\centering
\includegraphics[scale=0.5]{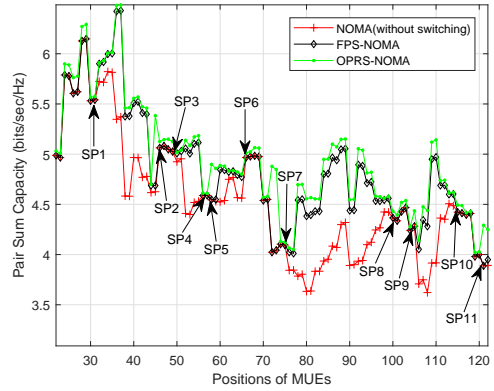}
\caption{Comparison of the proposed OPRS-NOMA with the FPS-NOMA and conventional NOMA in terms of PSC when mobile users switch roles multiple times.}
\label{Figure:4}
\end{figure}

The performance of proposed OPRS-NOMA is compared with conventional OMA and NOMA with fixed power and no role switching, and shown in Fig. \ref{Figure:3}. Switching positions can be seen in Fig. \ref{Figure:3}(a) to show the pre/post role switching gain. In addition, the comparison of PSC gains achieved by the OPRS-NOMA and FPS-NOMA between the two switching positions is shown in Fig. \ref{Figure:3}(b). Finally, to show the significant gains of OPRS-NOMA as compare to the FPS-NOMA, random behaviour of mobile users with multiple times role switching is presented in Fig. \ref{Figure:4}.

\section{Conclusions}
In this letter, a role switching and power allocation technique, OPRS-NOMA, is proposed for mobile users under RWP mobility model in NOMA. OPRS-NOMA overcomes the NPVP problem of mobile NOMA users by switching the roles on the basis of their channel gains when they come very close or cross each other. Power allocation is also updated based on the locations of mobile users. As performance measures, per user capacity and pair sum capacity are obtained by simulations. It can be seen that the proposed OPRS-NOMA outperforms the conventional NOMA and OMA with and without role switching.\\
In future, the performance gain in terms of capacity and BER will be discussed in detail. Moreover, addressing the user pairing problems and power allocation issues for multiple mobile users in NOMA are some interesting future research directions.
\ifCLASSOPTIONcaptionsoff
\fi

\end{document}